\documentclass[aps,pra,twocolumn,groupedaddress,floatfix,showpacs,nofootinbib]{revtex4-1}

\usepackage{float}
\usepackage[pdftex]{graphicx,color}
\usepackage{placeins}

\begin{document}

\title{Efficient Production of Large $^{39}$K Bose-Einstein Condensates}

\author{Robert L. D. Campbell, Robert P. Smith, Naaman Tammuz, Scott Beattie, Stuart Moulder, and Zoran Hadzibabic}

\affiliation{Cavendish Laboratory, University of Cambridge, J. J. Thomson Avenue, Cambridge CB3 0HE, United Kingdom}

\date{\today}

\begin{abstract}
We describe an experimental setup and the cooling procedure for producing $^{39}$K Bose-Einstein condensates of over $4\times10^5$ atoms. Condensation is achieved via a combination of sympathetic cooling with $^{87}$Rb in a quadrupole-Ioffe-configuration (QUIC) magnetic trap, and direct evaporation in a large volume crossed optical dipole trap, where we exploit the broad Feshbach resonance at 402$\,$G to tune the $^{39}$K interactions from weak and attractive to strong and repulsive.  In the same apparatus we create quasi-pure $^{87}$Rb condensates of over $8\times10^5$ atoms.
\end{abstract}


\maketitle

\section{Introduction}
\label{sec:intro}

Quantum degenerate atomic gases are widely used for studies of fundamental many-body physics in a highly controllable experimental setting~\cite{Bloch2008}, as well as for atom interferometry and precision measurements~\cite{Cronin2009}. The appeal of these systems primarily stems from the possibility to engineer both their geometry and interaction properties using the precision tools of atomic physics, such as optical lattice potentials~\cite{Morsch2006} and magnetically tunable Feshbach scattering resonances~\cite{Tiesinga1993, Inouye1998,Chin2010}. The primary goal of many-body research with ultra-cold atoms is to study universal collective quantum phenomena, which should in principle not depend on the details of the atomic isotopes used in the experiments, but only on their generic properties such as spin and interaction strength. In practice, however, the experimental flexibility in quantum-engineering of interesting many-body scenarios does often depend on the details of the atomic structure. In particular, the Feshbach resonances which can be used to tune the interactions in the gas are widely species dependent~\cite{Chin2010}. In some cases they occur at impractically high magnetic fields, while in others they are too narrow to allow fine tuning of the interactions.

Recently, through a series of pioneering experiments at LENS~\cite{Roati2007, Roati2007b, Fattori2008}, the bosonic $^{39}$K isotope has emerged as an extremely promising atomic species for both many-body research with tunable interactions~\cite{Roati2007b} and atom interferometry~\cite{Fattori2008}. This isotope features an extremely convenient $52\,$G broad Feshbach resonance centered at a moderate magnetic field of about $402\,$G~\cite{DErrico2007}. Unfortunately, $^{39}$K is also a comparatively difficult species to cool into quantum degeneracy. Firstly, the unresolved hyperfine structure in the excited electronic state limits the effectiveness of laser cooling methods, and thus adversely affects the initial conditions for evaporative cooling in conservative magnetic or optical traps. Secondly, at zero magnetic field the triplet $s$-wave scattering length is small and negative. This makes it additionally difficult to achieve the high elastic collision rates needed for evaporative cooling, and ultimately prevents Bose-Einstein condensation in a magnetic trap, since the condensate is unstable against collapse in the presence of attractive interactions~\cite{Gerton2000}. Optical trapping and Feshbach tuning of the scattering length to a large positive value must thus be employed already during the cooling stage of the experiment. The practical difficulty with this is that most standard optical dipole traps can be used to effectively trap atoms only at temperatures which are more than an order of magnitude lower than those achieved in laser cooling of $^{39}$K. So far, condensation of $^{39}$K has been achieved in only one group~\cite{Roati2007}.  Initially, condensates of $3\times10^4$ atoms were created, with the number later being improved to up to $10^5$ atoms~\cite{Zaccanti}.

Here we describe our experimental apparatus to produce a $^{39}$K Bose-Einstein condensate (BEC), and discuss the various optimization steps which allowed us to increase the number of condensed atoms to more than $4 \times 10^5$. As in~\cite{Roati2007}, we employ sympathetic cooling of $^{39}$K atoms immersed in a cloud of evaporatively cooled $^{87}$Rb atoms to reach a temperature suitable for effective optical trapping (in our case $\sim5\,\mu$K).  We achieve almost lossless sympathetic cooling in a QUIC magnetic trap~\cite{Esslinger1998}, increasing the phase space density of the $^{39}$K cloud by five orders of magnitude while losing less than half of the atoms.  The final cooling stage is performed in a large volume crossed dipole trap (CDT).  Employing the broad Feshbach resonance in the $|F, m_F \rangle = |1,1\rangle$ hyperfine state, we perform forced evaporative cooling of $^{39}$K at a scattering length of $5\,$nm. We cross the condensation temperature with $13\times10^5$ atoms and produce quasi-pure BECs containing $4\times10^5$ atoms, and having a lifetime longer than 10$\,$s.

The paper is divided into six sections. In Sec.~\ref{sec:experiment} we outline our experimental setup. In Sec.~\ref{sec:mot} we provide details of the the laser cooling stage of the experiment, and in Sec.~\ref{sec:quic} we describe the process of sympathetic cooling in the QUIC trap.  In Sec.~\ref{sec:cdt} we discuss the relative merits of sympathetic cooling and direct evaporation of optically trapped $^{39}$K, and find the latter favorable for the production of large degenerate samples.  We conclude with a summary of our results.

\begin{figure*}
\centering
\includegraphics[width=\textwidth]{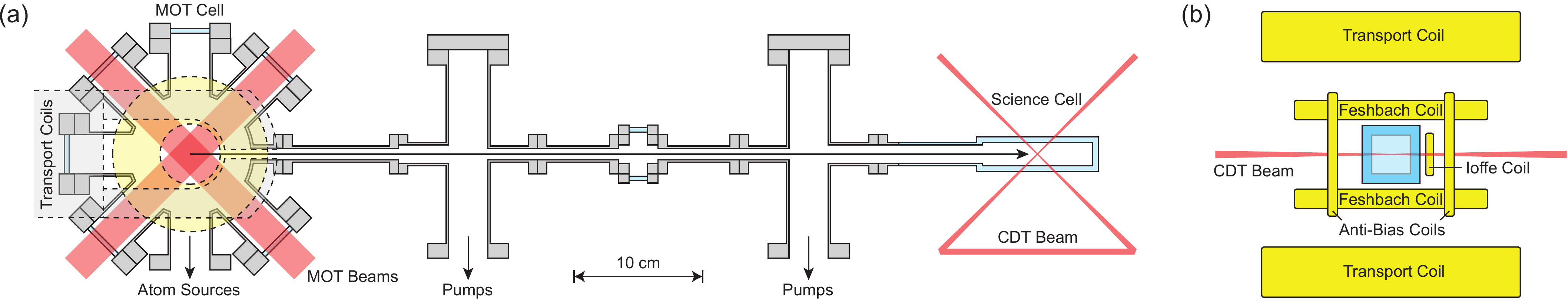}
\caption{Outline of the experimental system. (a) Horizontal cross-section through the vacuum chamber. Atoms are transported from the MOT cell to the science cell by the transport coils, mounted on a mechanical translation stage. (b) Region around the science cell as viewed down the transportation axis, showing the geometry of the coils used to create the QUIC trap and to access Feshbach resonances.}
\label{fig:Vacuum}
\end{figure*}

\section{Experimental Setup}
\label{sec:experiment}

\subsection{Vacuum system}

An overview of our experimental setup is shown in Figure \ref{fig:Vacuum}(a).  The system is divided into two ultra-high vacuum regions: the magneto-optical trap (MOT) region where laser cooling takes place, and the lower pressure ``science cell" in which BECs are made via sympathetic and evaporative cooling. Our MOT cell consists of a steel chamber fitted with eight anti-reflection coated windows (32$\,$mm viewable diameter).  The quartz science cell has a 30$\,$mm square outer cross-section, with wall thickness of 5$\,$mm.  The two chambers are separated by 60$\,$cm, and the differential pumping tubes connecting them have inner diameters of 10-16$\,$mm, designed to provide a differential pressure ratio between the two regions of $\sim\,$500. The ultra-high vacuum in the system is primarily maintained by two 55$\,$L$\,$s$^{-1}$ ion pumps, with additional pumping provided by a titanium sublimation pump located near the science cell.  

The rubidium and potassium vapor pressures in the MOT cell are maintained by periodically releasing atoms from commercial getters.  Our MOT loading times under typical experimental conditions are of the order of several seconds, while the lifetime of magnetically trapped (single-species) clouds in the science cell is $\sim300\,$s. The discrepancy between the observed and the designed differential pressure ratio is mainly due to local outgassing limiting the pressure reached in the science cell. 

\subsection{Laser Systems}
Four laser frequencies are required for the two-species MOT, two near 780$\,$nm for $^{87}$Rb and two near 767$\,$nm for $^{39}$K.  The rubidium cooling light close to the $|F=~2\rangle \rightarrow |F' = 3\rangle$ D$_2$ cycling transition is generated by an external cavity diode laser (ECDL) feeding a tapered amplifier (TA), while the weaker repumping light on the $|F = 1\rangle \rightarrow |F' = 2\rangle$ repumping line is produced by a separate ECDL.  In the case of $^{39}$K the excited state hyperfine structure is unresolved, so the demand on repumping power is higher. We split the output of a single ECDL into two beams of approximately equal power, shift their frequencies independently with acousto-optic modulators (AOMs), and then recombine the beams before amplifying them in the same TA.

All the cooling and repumping light for the two-species MOT is transferred to the experiment via a two- to six-way fibre port cluster.  The rubidium cooling light and the potassium light are combined on a polarizing beam splitter cube and then passed through a dichroic half-waveplate to align their polarizations.  This combined light is coupled into one of the input ports of the fibre cluster, while the $^{87}$Rb repumping light is coupled into the other.  Each cluster output provides one of the six four-frequency MOT beams, and zero order quarter waveplates are used to provide the required circular polarizations for both species.  For $^{87}$Rb, a total of 130$\,$mW of cooling light and 10$\,$mW of repumping light reaches the atoms in the MOT chamber. For $^{39}$K, the total laser power reaching the atoms is 310$\,$mW, and the balance between cooling and repumping light can be tuned by slight changes to the powers of the two beams seeding the same TA.  We use large collimated MOT beams, with a 1/$e^2$ diameter of $\sim$$\,3\,$cm, which essentially never require realignment.  The additional light required for optical pumping and imaging is split off from the ECDL beams prior to amplification in the TAs.

\subsection{Magnetic coils}
The same pair of transport coils shown in Fig. \ref{fig:Vacuum} provides the quadrupole field used for the MOT, for magnetic transport, and as part of the QUIC trap setup. They are wound from 4$\,$mm diameter hollow copper tube to allow water-cooling, and are mounted on a mechanical translation stage.  The coils provide a field gradient of 160$\,$G$\,$cm$^{-1}$ along the strong (vertical) axis for a maximum current of 200$\,$A.

The QUIC trap is produced with an additional Ioffe coil perpendicular to the transport coils, providing curvature along its axis, and a pair of anti-bias coils, providing a field offset along the same direction (see Fig. \ref{fig:Vacuum}(b)). The Ioffe coil consists of 18 turns of 1 mm copper wire, and has a clear inner diameter of 10 mm. The field it produces at the atoms can be modeled assuming that all turns are of radius 8.5$\,$mm and situated 18$\,$mm away from the center of the quadrupole field. It is mounted on a water-cooled copper block with a through-hole to allow imaging along the axis of the coil.  The rectangular anti-bias coils provide a field of 1.45$\,$G$\,$A$^{-1}$.  In the trap steady state 160$\,$A is run through the transport coils, 20$\,$A through the Ioffe coil, and 4$\,$A through the anti-bias coils. This results in trapping frequencies of $\omega_z/2\pi=14\,(21)\,$Hz axially and $\omega_r/2\pi=43\,(64)\,$Hz radially with respect to the Ioffe coil axis for rubidium (potassium) in the $|2,2\rangle$ ground state, and a bias field of 5$\,$G.

To exploit the broad Feshbach resonance in the $^{39}$K $|1,1\rangle$ state close to 402$\,$G, we use a pair of stationary coils close to the Helmholtz configuration.  At a current of 200$\,$A, the coil pair produces a uniform field of 525$\,$G at the position of the atoms.

\subsection{Crossed dipole trap}

Our crossed dipole trap is produced by a 10$\,$W ytterbium fibre laser with a wavelength of $\lambda =$ 1070$\,$nm.  A single laser beam is arranged in a ``bow-tie" configuration, as shown in Fig. \ref{fig:Vacuum}(a).  The beam is focussed by a concave/convex pair of lenses mounted in a cage system, allowing simple tuning of the beam waist between 100 and $160\,\mu$m. After passing through the vacuum system the beam is collimated, reflected and focussed back to cross itself perpendicularly with the same waist.  The depth of the optical trapping potential can be controlled over two orders of magnitude by a servo loop consisting of a photodiode, a PID controller, and an AOM.   A maximum power of 7.5$\,$W per beam makes it to the atoms, though we normally limit this to 7$\,$W to ensure power stability.

\subsection{Absorption imaging}

We image the atoms with resonant circularly polarized $|F=2\rangle\rightarrow|F'=3\rangle$ light.  Assuming perfect absorption cross-section typically underestimates atom numbers, mainly due to imperfections in the imaging polarization. We calibrate the cross-section by observing the critical temperature $T_{\text{c}}$ of a weakly interacting gas for $^{39}$K, while accounting for the $T_{\text{c}}$ interaction shift for $^{87}$Rb~\cite{Gerbier2004}.  The shifts in $T_{\text{c}}$ due to finite size~\cite{Dalfovo1999} and trap anharmonicity are less than 1\% and 3\% respectively.  We observe a reduction in cross-section by a factor of 1.9 for $^{39}$K, and 1.5 for $^{87}$Rb.  These values lie inside the range of factors typically reported by other groups~\cite{Gerbier2004,Sanner2010}.

\section{Two-species MOT}
\label{sec:mot}

We operate our two-species MOT with a vertical magnetic field gradient of 8$\,$G$\,$cm$^{-1}$. For $^{87}$Rb, the cooling light is detuned by -3.1$\,\Gamma$ from the cycling transition, where $\Gamma$ is the natural linewidth $\Gamma=\Gamma_{\text{Rb}}\approx\Gamma_{\text{K}}\approx2\pi\times6\,$MHz, while the repumping light is tuned close to resonance with the $|F = 1\rangle \rightarrow |F' = 2\rangle$ transition.  For $^{39}$K, a wide velocity capture range is achieved by detuning both laser frequencies below the entire (unresolved) upper-state hyperfine manifold.  During the MOT loading phase we use detunings of -7.6 and -4.6$\,\Gamma$ from the potassium $|F = 2\rangle \rightarrow |F' = 3\rangle$ and $|F = 1\rangle \rightarrow |F' = 2\rangle$ cooling and repumping transitions respectively. However, during the last 40$\,$ms of the MOT stage we decrease the two detunings to -1.5 and -4.1$\,\Gamma$ respectively, and also slightly reduce the repumping intensity.  This results in lower temperatures without any noticeable drop in atom numbers~\cite{Fort1998}.

Under standard operating conditions, the atom numbers in our single-species $^{87}$Rb and $^{39}$K MOTs saturate at about $5\times10^9$ and $1\times10^9$ respectively, after about 10$\,$s of loading.  However, for our sympathetic cooling experiments we require a higher ratio  of rubidium to potassium atom numbers (see Section \ref{sec:quic}). Hence, in practice we turn on the potassium light only once the rubidium MOT has almost saturated and vary the potassium loading time between 0.5 and 2$\,$s in order to obtain the desired number of potassium atoms in the QUIC magnetic trap.  The MOT stage is followed by 4$\,$ms of optical molasses, during which the $^{87}$Rb atoms reach $\sim50\,\mu$K while the $^{39}$K atoms reach several hundred $\mu$K.

\section{Sympathetic cooling in QUIC magnetic trap}
\label{sec:quic}

\subsection{Magnetic trapping and transport}

For magnetic trapping, both species are optically pumped into their stretched $|F=2,m_F=2\rangle$ hyperfine ground states, with circularly polarized $|F=2\rangle\rightarrow|F'=2\rangle$ pumping light and $|F=1\rangle\rightarrow|F'=2\rangle$ repumping light. Optical pumping efficiency is $\sim\,$80\% for $^{87}$Rb and $\sim\,$65\% for $^{39}$K.

To capture the atoms the quadrupole coils are abruptly switched on at $40\,$G$\,$cm$^{-1}$, and then ramped up to 160$\,$G$\,$cm$^{-1}$ in 50$\,$ms. We transport the atoms to the science cell in 2$\,$s, during which time atoms (mainly potassium) with an energy greater than $\sim2.7\,$mK are removed by the walls of the vacuum system.

\subsection{Choice of the trapping potential for sympathetic cooling}

After transportation to the science cell the mixture of $^{87}$Rb and $^{39}$K atoms is held in the tightly confining quadrupole potential generated by the transport coils. With $^{87}$Rb alone it is possible to evaporate directly in the quadrupole trap before loading into the CDT for continued forced evaporation.  This procedure enables production of a BEC of $5\times10^5$ $^{87}$Rb atoms in less than ten seconds. However, this approach is not successful when trying to achieve sympathetic cooling of $^{39}$K. Potassium is not effectively cooled for such short evaporation sweeps due to the relatively small K-Rb interspecies background scattering length, $a_{\text{KRb}}\simeq28\,a_0$\cite{Ferlaino2006}, where $a_0$ is the Bohr radius.  For comparison, the rubidium intraspecies background scattering length is $a_{\text{Rb}}\simeq99\,a_0$\cite{Kempen2002}. Long evaporation sweeps in the quadrupole trap are prohibited by inelastic collisions with $^{87}$Rb atoms in the $|2,1\rangle$ state.  These are created by Majorana spin-flips, despite the presence of the CDT offset from the magnetic zero.

We avoid this issue by transferring the atoms to a QUIC trap, which has a finite magnetic field minimum. This also enables a microwave field to be applied that is only resonant with atoms in the $|2,1\rangle$ state, efficiently removing the few that are produced during evaporation. This trap is also found to be effective for single-species cooling of $^{87}$Rb, ultimately leading to the production of condensates of up to $8\times10^5$ atoms in the CDT. We ramp up the Ioffe and anti-bias coils to load $4\times10^8$ rubidium atoms and up to $6\times10^7$ potassium atoms into the QUIC trap~\cite{Footnote1}.

\subsection{Sympathetic cooling results}

\begin{figure}
{\includegraphics*[width=0.85\columnwidth]{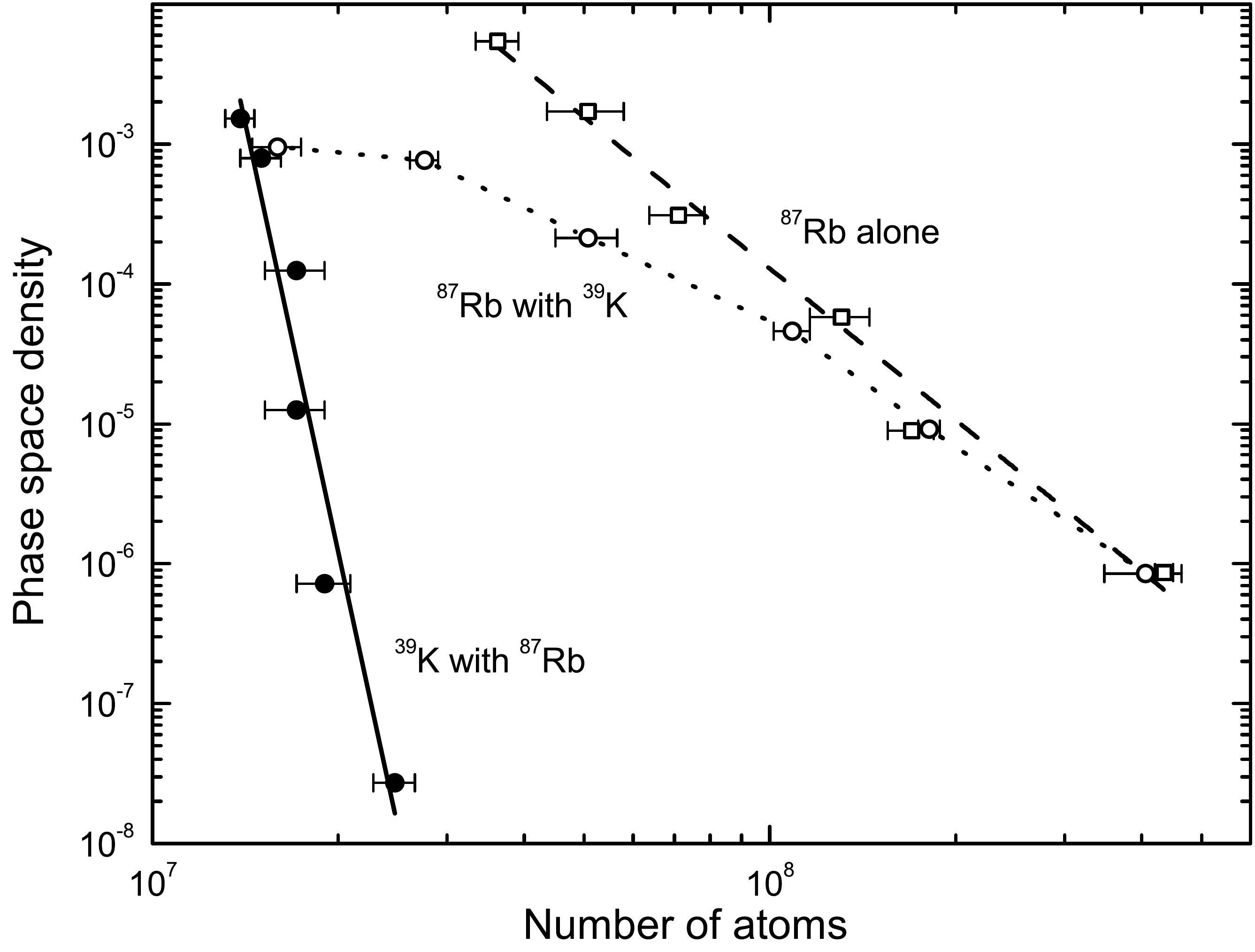}
\put(-230,155){(a)}}
{\includegraphics*[width=0.85\columnwidth]{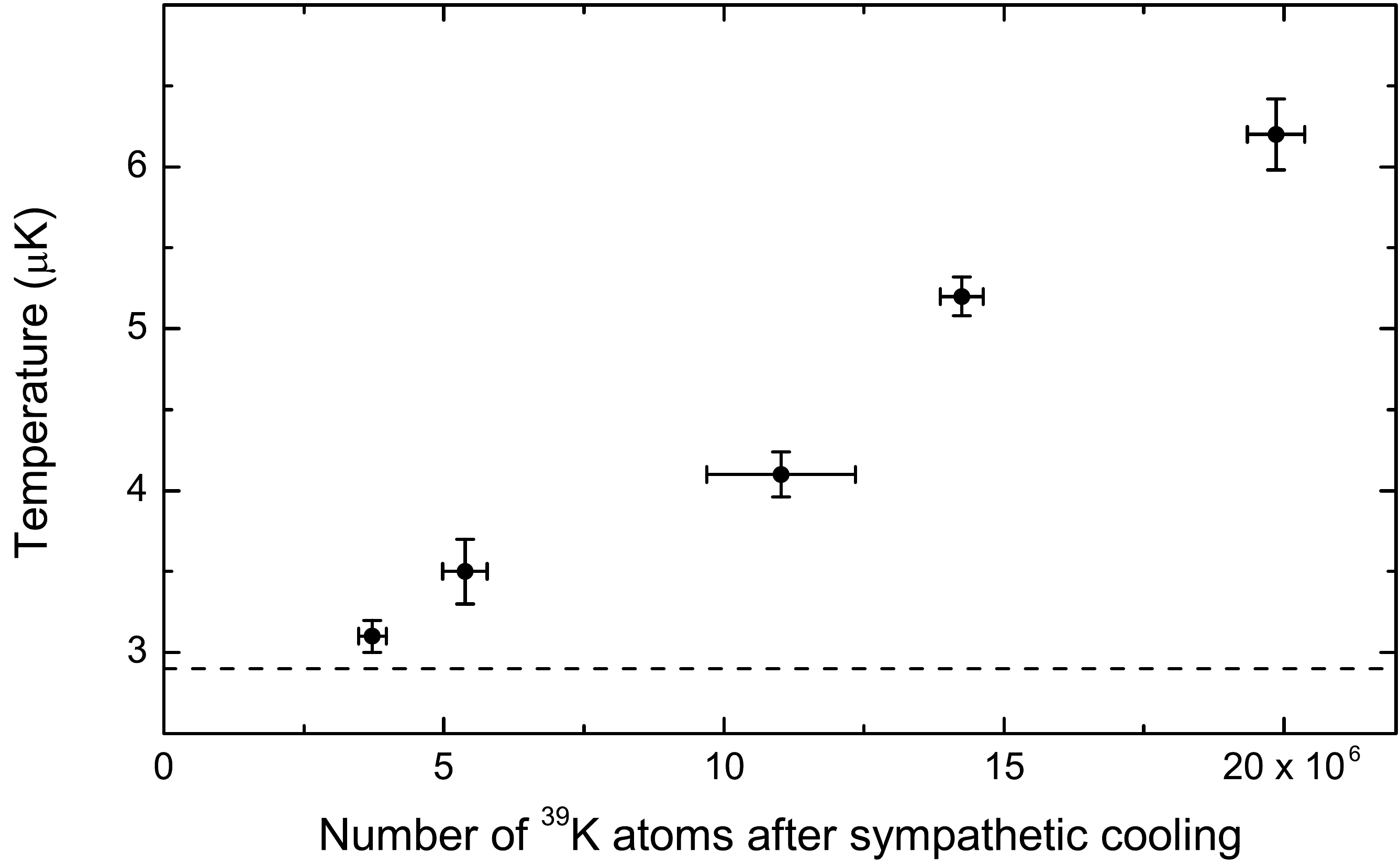}
\put(-230,130){(b)}}
\caption{Evaporative and sympathetic cooling in the QUIC trap. (a) Cooling trajectories in single- (squares) and two-species (circles) experiments. Open and solid markers represent $^{87}$Rb and $^{39}$K respectively. The two fits show the difference between the steep trajectory for sympathetic cooling (solid line, $\gamma_{\text{sym}}=20$) and that for efficient evaporative cooling (dashed line, $\gamma=3.6$). Vertical error bars lie within the markers. (b) Final temperature reached by sympathetic cooling for different sizes of the $^{39}$K load.  The dashed line shows the temperature reached by $^{87}$Rb in the absence of $^{39}$K.}
\label{fig:quicplot}
\end{figure}

In any sympathetic cooling experiment there is a natural tradeoff between the size of the sympathetically cooled load and the lowest achievable temperature.  In our experiments we aim to cool the largest possible $^{39}$K sample to a temperature of about 5$\,\mu$K. This goal is motivated by two factors: Firstly, in our QUIC trap the difference between the gravitational sags, $g/\omega_r^2$, for rubidium and potassium is 73$\,\mu$m.  This corresponds to the (vertical) thermal radii for the two species, $R_{\text{T}}^{\text{Rb}}\approx R_{\text{T}}^{\text{K}}$, at a temperature of about 4$\,\mu$K, where $R_{\text{T}}=\sqrt{k_{\text{B}}T/m\omega_r^2}$. We can therefore anticipate that, even for very small potassium loads, sympathetic cooling becomes ineffecient below approximately that temperature, due to the reduced spatial overlap of the clouds~\cite{Delannoy2001}. Secondly, as discussed in the following section, this temperature is closely matched with the optimal temperature for loading the largest possible potassium samples into our optical trap.

The results of our sympathetic cooling experiments are summarized in Fig. \ref{fig:quicplot}. In Fig. \ref{fig:quicplot}(a) we first characterize our rubidium cooling reservoir. Allowing for the fact that for the same trap depth the temperature in the presence of a potassium load is inevitably slightly higher, here we set the target temperature to about 3$\,\mu$K. We employ a 56$\,$s forced evaporation on the microwave $|2,2\rangle\rightarrow|1,1\rangle$ $^{87}$Rb hyperfine transition, using an exponential sweep from 6873.68 to 6847.24$\,$MHz with a 14$\,$s time constant. The evaporation of $^{87}$Rb alone is characterized by a single parameter $\gamma=-d\ln(D)/d\ln(N)$, where $D$ is the phase space density in the centre of the trap and $N$ is the atom number. We observe very efficient cooling with $\gamma=3.6$~\cite{Ketterle1996}.  The long sweep is required to allow thermalization between $^{39}$K and $^{87}$Rb due to the small value of $a_{\text{KRb}}$, while in rubidium-only experiments we observe almost identical efficiency if we shorten the evaporation time to 32$\,$s. Note that in these experiments we continually sweep a separate microwave source between 6838 and 6844$\,$MHz.  Due to the 5$\,$G field offset at the minimum of our QUIC trap, this removes any unwanted $|2,1\rangle$ $^{87}$Rb atoms without affecting the atoms in the $|2,2\rangle$ state.

In the same panel we show the evolution of the phase space density of a typical sympathetically cooled $^{39}$K cloud used in our experiments.  The almost vertical cooling trajectory is the hallmark of effective sympathetic cooling. We cool the $^{39}$K cloud from 200 to $5\,\mu$K, while the atom number is reduced by less than a factor of two, approximately from $25\times10^6$ to $14\times10^6$.  In analogy with conventional evaporative cooling we can express the efficiency of this process through an effective $\gamma_{\text{sym}}\approx20$. 

Finally, we also show the effect of the $^{39}$K load on the cooling trajectory of the $^{87}$Rb cloud. The path starts to deviate from that of the rubidium-only trajectory when the potassium number becomes a significant fraction of the rubidium number ($N_{\text{Rb}}/N_{\text{K}}<10$), and bends to almost meet the potassium cooling path at the end of the evaporation sweep.  The convergence of these two trajectories suggests that with this potassium load we approach the limits of the effectiveness of sympathetic cooling. Further evaporation past the point where the heat capacities of the two species become comparable will have little benefit for the cooling of potassium.  

Fig. \ref{fig:quicplot}(b) shows the final temperature achieved for various potassium numbers in the QUIC, showing the trend towards the temperature of 2.9$\,\mu$K achieved by rubidium (when cooled alone) for low $N_{\text{K}}$.  The quantity of $^{39}$K that we choose to load into the MOT is determined by the number of atoms that can later be loaded into the CDT after cooling in the QUIC. This depends on the temperature reached after sympathetic cooling, as discussed below. A 20\% reduction in $N_{\text{K}}$ over the evaporation period would be accounted for by the trap lifetime, so there is evidence for some inelastic processes, but the effect is not significant.

\section{Bose Einstein Condensation in Optical Trap}
\label{sec:cdt}

\subsection{Loading of atoms into the CDT} 

The CDT laser power is ramped up to 7$\,$W in 1$\,$s, and the currents in the three QUIC coils are ramped down by a factor of two in a further 1$\,$s. The quadrupole and Ioffe coils are then switched off abruptly, leaving the atoms trapped in the CDT which has a depth at this point of $\sim\,$30$\,\mu$K for $^{39}$K ($\sim\,$35$\,\mu$K for $^{87}$Rb). 

\begin{figure}
\centering
\includegraphics[width=0.85\columnwidth]{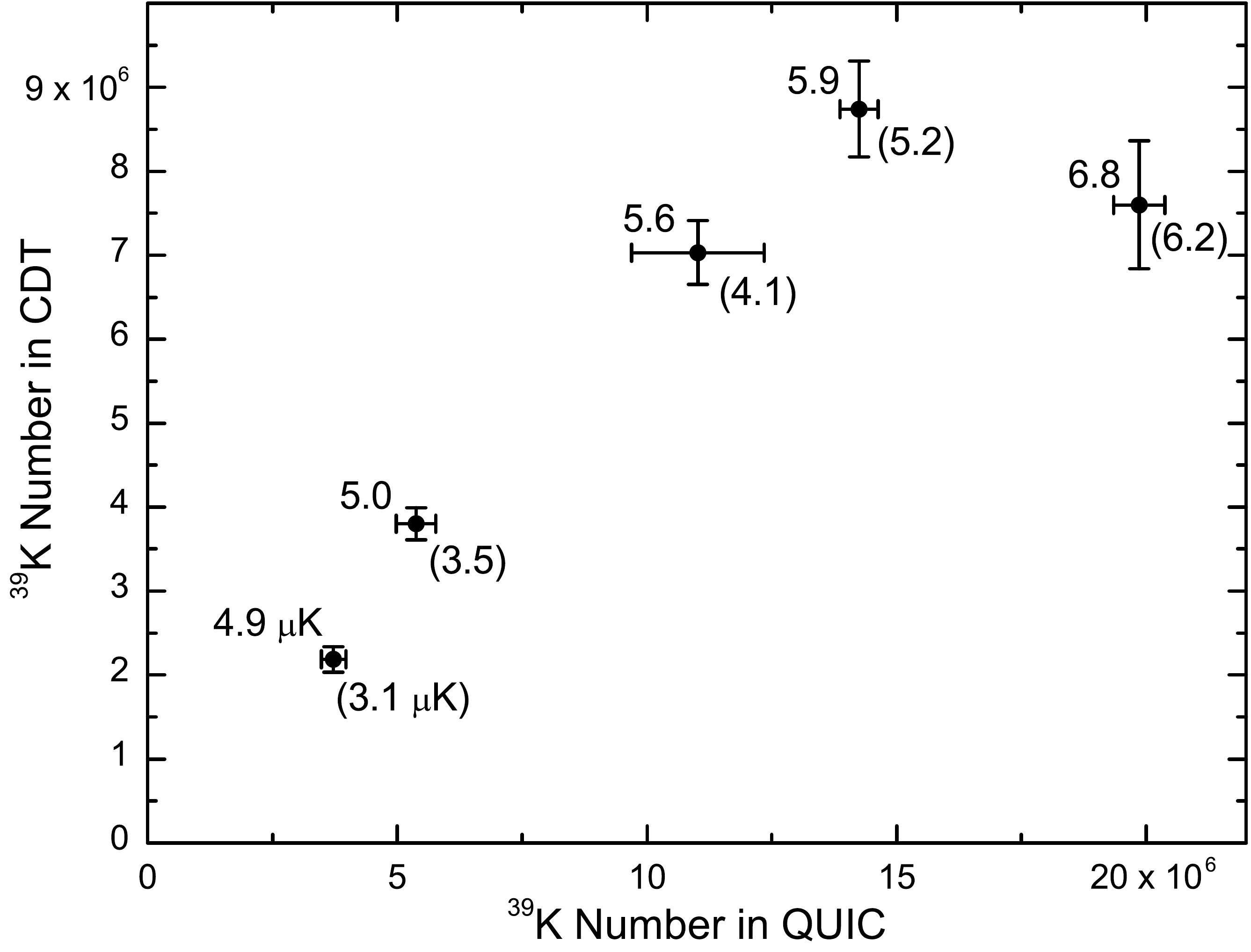}
\caption{Transfer efficiency from QUIC to CDT. The numbers at each point are the temperatures in the CDT (QUIC).}
\label{fig:transfer}
\end{figure}

The number of $^{39}$K atoms that are transferred to the CDT as a function of the number in the QUIC is shown in Fig. \ref{fig:transfer}. We observe a constant transfer efficiency of $\sim\,$60$\,$\% for samples with a temperature of up to 6$\,\mu$K, equivalent to $\eta\,$$\sim\,$5, where $\eta$ is the ratio of the trap depth to the temperature of the atoms.  This constant factor is due to imperfect overlap between the QUIC and CDT potentials.  This is challenging to overcome due to the mismatch in geometries between the two traps, though the large beam waist of the CDT helps~\cite{Footnote2}. For comparison, loading efficiency was tested for a 100$\,\mu$m beam waist and found to be limited to $\sim\,$20\%, saturating at around a third of the available CDT power. Increasing the waist beyond 140$\,\mu$m is currently not an option without decreasing the available trap depth as we are working at our maximum laser power. 

In practice, we load $8\times10^6$ $^{39}$K atoms at 6$\,\mu$K into the CDT from $13\times10^6$ left at the end of the QUIC.  Rubidium transfer efficiency is slightly lower because the CDT alignment is optimized for potassium.

\subsection{Cooling schemes}

\begin{figure}[t]
\centering
\includegraphics[width=0.85\columnwidth]{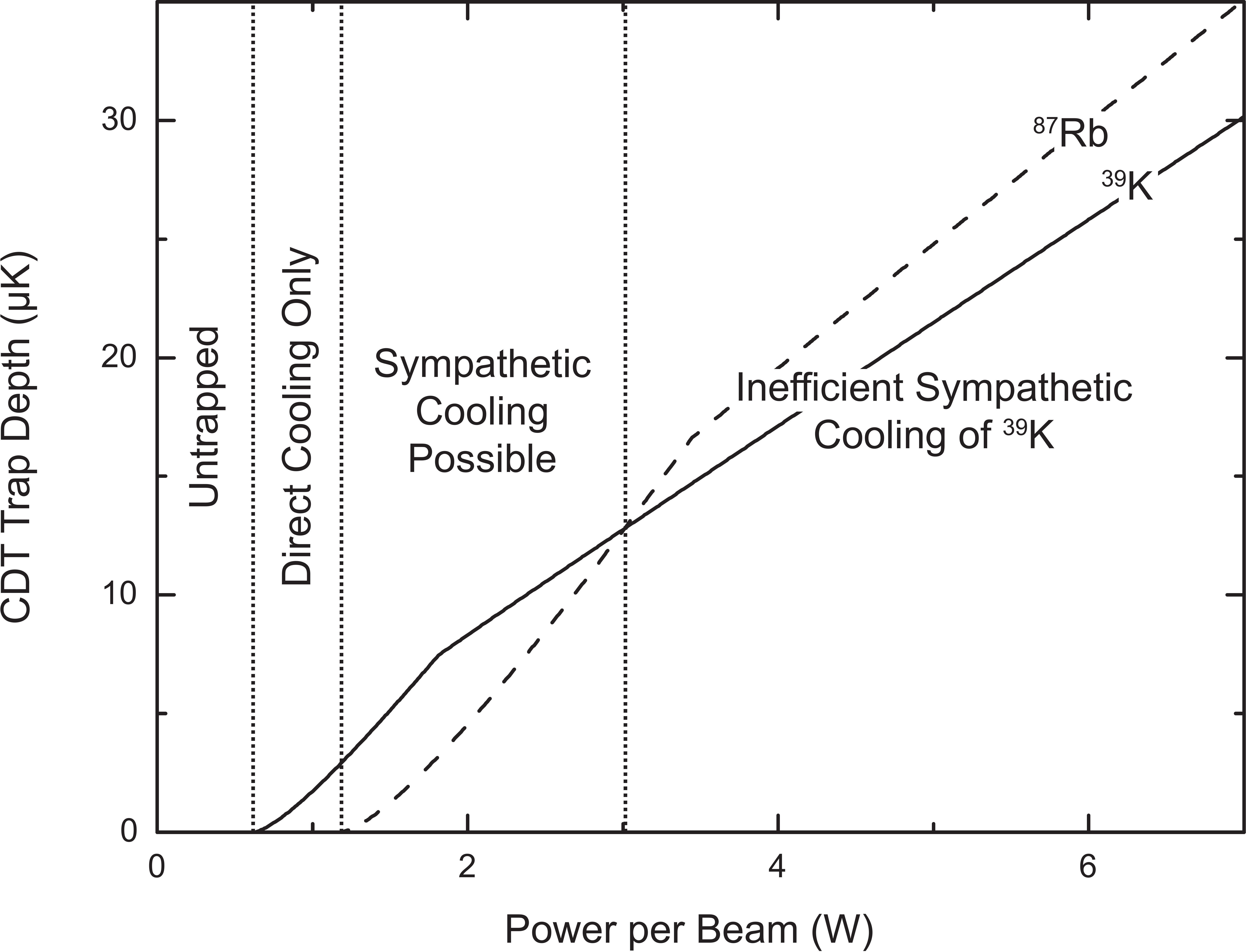}
\caption{CDT depth versus power for both $^{87}$Rb (dashed line) and $^{39}$K (solid line) for a beam waist of $140\,\mu$m. There are two regimes where different mechanisms limit the trap depth. At high power the atoms are least confined along the beams whereas at lower power they are less confined vertically, due to the effect of gravity. There is a clear kink at the crossover. The marked zones illustrate the possibility of sympathetically cooling $^{39}$K with $^{87}$Rb for low trap depths and the necessity to abandon the scheme for deeper traps.}
\label{fig:cdt}
\end{figure}

Once the atoms are loaded into the CDT they are no longer in a state-dependent potential, opening up options for the final stages of cooling.  We can choose to carry on sympathetically cooling the potassium until all the rubidium is depleted before directly cooling the potassium with the aid of a Feshbach resonance, or we can simply directly cool the potassium all the way.  The former method works very well for small optical trap depths~\cite{Roati2007}.  However, this approach is unsuitable for cooling large samples in deep traps.  The difference between the two regimes is illustrated in Fig. \ref{fig:cdt}.  The relative trap depths for $^{87}$Rb and $^{39}$K depend on the competition of two factors: the strength of the electric dipole potential for the two species, and the deformation of the potential by gravity.  Both effects are larger for rubidium, the former due to the smaller detuning of the CDT beam from the $^{87}$Rb D line, the latter because of its higher mass.  At high laser intensities the effect of gravity is negligible and atoms are preferentially lost along the beams of the CDT such that the trap depth is higher for rubidium.  Lowering the intensity causes the gravitational force to distort the Gaussian optical potential to the point at which atoms leave from the bottom of the trap. Eventually this effect causes the trap depth for rubidium to fall below that of potassium, and only then will rubidium act effectively as a coolant for efficient sympathetic cooling.

It is possible to extend the temperature range over which sympathetic cooling of $^{39}$K is effective. The trap depth for our 1070$\,$nm beam at the crossover to the sympathetic region, $U_{\text{cross}}$, has been found numerically to scale as $U_{\text{cross}}/w=0.091\,\mu$K$\,\mu$m$^{-1}$, where $w$ is the beam waist. So, for example, to extend this range up to 30$\,\mu$K would involve increasing the beam waist to 330$\,\mu$m, requiring 40$\,$W of trapping light. As discussed above, this is currently not viable for us, and we therefore choose to evaporate $^{39}$K on its own in the CDT, proceeding as follows.

\subsection{Direct Evaporation of $^{39}$K in the CDT}

\begin{figure}[t]
\centering
\includegraphics[width=0.95\columnwidth]{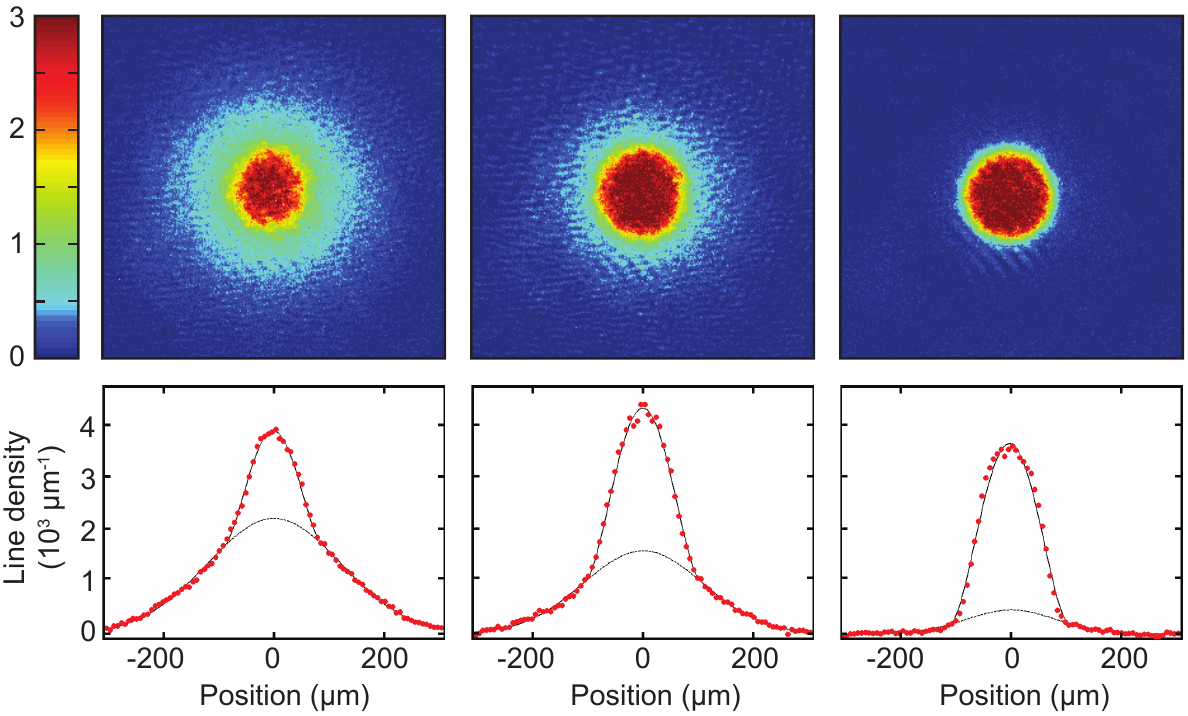}
\caption{Condensation of $^{39}$K atoms.  From left to right the number of atoms in the BEC varies from $1.5\times10^5$ to $4.2\times10^5$, while the condensate fraction grows from 20\% to 80\%. The bottom panels show the line densities of the clouds, obtained by integrating the images along the vertical direction. Images are taken after 20$\,$ms time-of-flight. The changing aspect ratio of the condensate reflects the increasing importance of the gravitational potential at low CDT powers.}
\label{fig:becs}
\end{figure}

The remaining $^{87}$Rb atoms are removed from the trap with a $100\,\mu$s resonant light pulse, in order to prevent any spin-exchange collisions during the state transfer of the $^{39}$K atoms from the $|2,2\rangle$ to the $|1,1\rangle$ state. The state transfer is accomplished by a 100$\,$ms Landau-Zener sweep of the bias magnetic field in the presence of 469$\,$MHz radiation.  The state transfer efficiency is larger than 90\%.

A uniform magnetic field is then ramped up over 5$\,$ms to the vicinity of the 52$\,$G-wide $^{39}$K Feshbach resonance centered at 402.5$\,$G \cite{DErrico2007,Zaccanti2009}. We typically use a field of 391$\,$G which gives a scattering length of $100\,a_0$.  We then continue forced evaporation by lowering the CDT power in a 6$\,$s sweep until a BEC is formed as shown in Figure \ref{fig:becs}. Using this procedure we can produce condensates of over $4\times10^5$ atoms with a lifetime of more than 10$\,$s. In order to image the atoms we first optically pump them back into the $|F=2\rangle$ state, before imaging on the $|F=2\rangle\rightarrow|F'=3\rangle$ transition. We can produce BECs by evaporation at scattering lengths of up to $250\,a_0$, but the largest and most stable condensates are produced at more moderate scattering lengths near 100$\,a_0$.

\section{Conclusion}
\label{sec:conclusion}

We have described our apparatus and procedure for producing $^{39}$K Bose-Einstein condensates using a combination of sympathetic cooling with $^{87}$Rb in a QUIC magnetic trap and direct evaporation in a large volume crossed dipole trap. We produce condensates of over $4\times 10^5$ atoms with a lifetime of 10$\,$s. For comparison, our $^{87}$Rb BECs produced in the same setup contain up to $8\times10^5$ atoms, when optimized for rubidium alone. The broad Feshbach resonance at 402.5$\,$G and low background scattering length make this is an ideal system to study the effects of tuning the atomic interaction strength on the properties of Bose gases. 

We have optimized the system for the available resources, but we do not observe any fundamental limits to improving the degenerate $^{39}$K samples further.  This could be achieved by increasing the size of the sympathetic cooling reservoir and the available CDT power. 

\FloatBarrier

\begin{acknowledgments}
We thank M. K\"{o}hl, M. Inguscio, G. Modugno, G. Roati, M. Zwierlein, and J. Dalibard for useful discussions, I. Gotlibovych, T. Schmidutz, P. Richman, N. Engelsen, and A. Dareau for experimental assistance, and A. Keshet for the Cicero experimental control software.  This work is supported by EPSRC (Grant Nos. EP/G026823/1 and EP/I010580/1).

\end{acknowledgments}

\providecommand{\noopsort}[1]{}\providecommand{\singleletter}[1]{#1}%


\begin{thebibliography}{10}
\bibitem{Footnote2}%
  \BibitemOpen
  \bibinfo {note} {Radially, $R_{\text{T}}^{\text{K}}\simeq80\,\mu$m for
  $^{39}$K at 5$\,\mu$K in the QUIC trap, meaning $\sim\,$95\% of the atoms lie
  within the effective $\sqrt{2}\times140$$\,\mu$m beam waist of the CDT at its
  intersection with the magnetic trap.}%
  \bibAnnoteFile{Stop}{Footnote2}%
\end{thebibliography}
\end{document}